\documentclass[prc,preprint,showpacs]{revtex4}
\usepackage{hyperref}
\usepackage{graphicx}
\usepackage{amssymb}
\usepackage{amsmath}

\begin{document}

\title{Nuclear liquid-gas phase transition studied with
  antisymmetrized molecular dynamics}
\author{Takuya Furuta and Akira Ono}
\affiliation{Department of Physics, Tohoku University, Sendai
980-8578, Japan}
\begin{abstract}
The nuclear liquid-gas phase transition of the system in ideal thermal
equilibrium is studied with antisymmetrized molecular dynamics.  The
time evolution of a many-nucleon system confined in a container is
solved for a long time to get a microcanonical ensemble of a given
energy and volume.  The temperature and the pressure are extracted
from this ensemble and the caloric curves are constructed.  The
present work is the first time that a microscopic dynamical model
which describes nuclear multifragmentation reactions well is directly
applied to get the nuclear caloric curve.  The obtained constant
pressure caloric curves clearly show the characteristic feature of the
liquid-gas phase transition, namely negative heat capacity
(backbending), which is expected for the phase transition in finite
systems.
\end{abstract}

\maketitle

\section{Introduction}
\label{INTRO}
Phase transition is an interesting phenomenon which appears in various
physical systems.  In nuclear systems with the excitation energies of
a few to ten MeV/nucleon, the existence of the liquid-gas phase
transition has been speculated based on the resemblance between the
equation of state (EOS) of homogeneous nuclear matter and the Van der
Waals EOS. However, the confirmation of the phase transition for
realistic nuclear systems requires more careful arguments.

First of all, any nuclear system accessible by experiments consists of
a finite number of nucleons.  Because of this, it may be generally
believed that the phase transition tends to be smeared out due to the
finite size effect, which is a correct statement for canonical
ensembles specified by temperature.  However, by studying
microcanonical ensembles with fixed energies, a quite prominent signal
of the phase transition is expected in finite systems, that is
backbending (or negative heat capacity) in caloric curves
\cite{grossPR,grossEP,grossPCCP}.  Many experimental and theoretical
works have been devoted to search for such a signal in various
physical systems
\cite{gobet1,gobet2,schmidt,pochodzalla,chomaz,agostinoNP,raduta,gulminelli}.

Another complexity comes from the fact that the excited systems for
the study of nuclear liquid-gas phase transition are produced in
dynamical processes of nuclear reactions in laboratories. In fact, it
is considered that the liquid-gas phase transition is somehow related
to the multifragmentation phenomenon which is observed in the various
nuclear reactions, such as in the relatively low energy region where
the neck or midrapidity component may be created in dissipative binary
reactions \cite{montoya,lukasik,lefort}, in higher energy central
collisions where clusters are produced copiously in expanding system
\cite{reisdorf}, and also in peripheral collisions where the excited
projectile-like fragment breaks up into pieces \cite{pochodzalla}.
Thus intensive studies have been done to search evidences of the phase
transition in these experimental data and some of the works have
reported that indications of the phase transition have been obtained
\cite{guptaExpEvi,pochodzalla,agostinoNP,agostinoFLCT,natowitz,elliott,trautmann}.
However, the conclusion has not come yet and much efforts are still
required. The difficulty is due to the nontrivial dynamical effects
contained in the experimental data.  Even though it may be true that
the part of the system is equilibrated in good approximation as is
indicated by the success of statistical models
\cite{randrupSM,bondorfSM,agostinoSM,grossPR}, the ambiguities enter
in the data analysis through the identification of the equilibrated
thermal source.  On the other hand, there are several models which can
predict the experimental observables without such ambiguities by
simulating the dynamics of these reactions microscopically
\cite{bertsch,cassing,aichelin,maruyama,onoPRL,onoPTP}.  By carefully
studying the time evolution of dynamical reactions, it may be possible
to discuss how well the thermalization is achieved in dynamical
collisions and how the results of dynamical reactions are related to
the thermal properties, such as phase transition. For this purpose,
the dynamical models should correctly describe the thermal properties
as well as dynamical reaction mechanisms. It can be verified by
applying these models directly to the ideal systems in thermal
equilibrium, though the attempts for this direction have not been done
sufficiently.

The aim of the present work is to demonstrate that the nuclear
liquid-gas phase transition in ideal thermal equilibrium can be
described by antisymmetrized molecular dynamics (AMD) which is a
microscopic dynamical model based on the degrees of freedom of
interacting nucleons \cite{onoPRL,onoPTP,onoReview}.  We utilize the
same AMD model that has been applied to nuclear collisions and
successfully reproduced various aspects of experimental data
\cite{onoPRL,onoPTP,onoCa,onoAu,onoWithFrance,onoReview,wada1998,wada2000,wada2004}.

To achieve this aim, the equilibrated systems are constructed and
their statistical properties are studied in the following way.
Firstly, we prepare the initial state putting $A$ nucleons ($N$
neutrons and $Z$ protons) arbitrarily within a fixed radius
$r_{\text{wall}}$, and solve the time evolution of the system by AMD
for a long time. In order to confine the system in the container of
volume $V=\frac{4}{3}\pi r_{\text{wall}}^3$, we introduce a certain
reflection process at the container wall. Then we regard the state at
each time as a sample of the statistical ensemble with energy $E$,
volume $V$ and particle number $A=N+Z$. It should be emphasized that
this ensemble is a microcanonical ensemble with a fixed energy $E$
because the initial energy $E$ is conserved through the AMD time
evolution. By extracting the statistical information (temperature $T$
and pressure $P$) from the ensembles, we can construct the caloric
curves. The existence of the liquid-gas phase transition can be
checked by finding a part with negative heat capacity (backbending) in
these caloric curves, which is the characteristic feature of first
order phase transition for finite
systems\cite{grossPR,grossEP,grossPCCP}. 

One of the advantages of the present study as a statistical model,
compared to conventional statistical models
\cite{grossPR,randrupSM,bondorfSM}, is that the calculation is based
on the nucleon-nucleon interaction, rather than the nuclear binding
energies and level densities given externally.  Another advantage is
that the interactions among fragment nuclei and nucleons are naturally
taken into account, while they are usually ignored in other
statistical models by the freeze-out assumption.

There have been several discussions on the question whether molecular
dynamics models can describe the statistical properties of nuclear
systems for which quantum and fermionic features are essential
\cite{ohnishiPRL,ohnishiAP,FMDLGpt,onoAMDMF,onoSts}.  It has been
shown that molecular dynamics can be consistent with the quantum and
fermionic caloric curve if the wave function is fully antisymmetrized
and an appropriate quantum branching process is taken into account
\cite{ohnishiPRL,ohnishiAP,onoAMDMF,onoSts}.

There are several works
\cite{dorso,peilert,ohnishiPRL,ohnishiAP,FMDLGpt,onoSts,sugawaVolkov,sugawaGogny}
that studied the nuclear liquid-gas phase transition using molecular
dynamics models and some of the works claimed that a clear signal of
phase transition, namely plateau or backbending in the caloric curves,
is obtained. However, the molecular dynamics models utilized in these
works have not been successfully applied to the dynamical
multifragmentation reactions.  Another problem is that the effective
interactions adopted in these works sometimes do not satisfy the
saturation property of infinite nuclear matter and none of them
succeeded in obtaining plateau or backbending with an appropriate
effective interaction so far.  Furthermore, these works have
calculated the constant volume caloric curves or those with a harmonic
oscillator confining potential but have not explicitly shown the
constant pressure caloric curves for which plateau or backbending is
expected most clearly.  On the other hand, our present study uses the
same framework of AMD (AMD/DS) that has already been utilized for
nuclear collision simulations \cite{onoWithFrance,onoReview} and
the Gogny force \cite{gogny} as the effective interaction between
nucleons, which satisfies the saturation property of infinite nuclear
matter.  We also draw the caloric curves at constant pressure.

This paper is organized as follows.  In Sec. \ref{FORCE}, we show
features of the Gogny force when it is applied to infinite nuclear
matter. In Sec.\ \ref{FRAMEWORK}, the framework of the AMD is
explained, which is used to calculate the time evolution of
many-nucleon system to create a microcanonical ensemble. The
reflection process to confine the system in a given volume is also
explained.  In Sec.\ \ref{CALC}, the method to extract the statistical
information (temperature $T$ and pressure $P$) from our calculations
is described. Then in Sec.\ \ref{RESULT}, the obtained caloric curves
with constant volume and also with constant pressure are shown. The
dependence of the caloric curves on the theoretical ambiguities is
also discussed.  Section \ref{SUMMARY} is devoted to a summary and
future perspectives.

\section{Effective interaction and the property of infinite nuclear matter}
\label{FORCE}
\begin{figure}
\begin{center}
\includegraphics[width=6cm,clip]{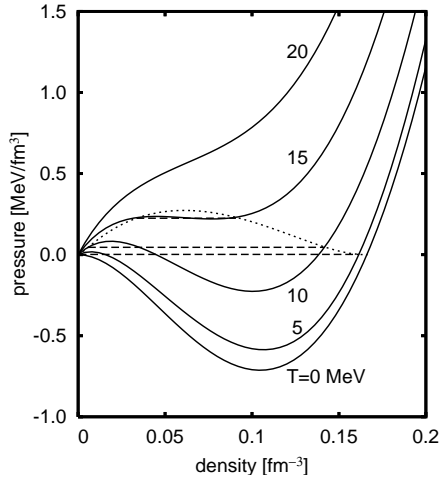}
\end{center}
\caption{The EOS of symmetric nuclear matter.  Each full line shows
  the pressure as a function of the density for the given temperature
  (0, 5, 10, 15 and 20 MeV). This EOS is obtained by adopting the
  Gogny force and assuming uniform density in the Hartree-Fock
  framework.  Below the dotted line, the liquid and gas phases coexist
  if nonuniform density is allowed.  The dashed lines are the EOS in
  this coexistence region obtained by the Maxwell construction.}
\label{fig:eos}
\end{figure}

In the present study, we adopt the Gogny force \cite{gogny} as the
effective nuclear interaction. The Gogny force is one of the most
successful effective interactions to reproduce the ground state
properties of nuclei in mean field theories.  The Gogny force
satisfies the saturation property of the nuclear matter with the
incompressibility $K=228$ MeV when it is applied to infinite nuclear
matter with the uniform Hartree-Fock approximation.  The saturation
property is essential for the discussion of the similarities and the
differences between finite and infinite nuclear systems.  The equation
of state (EOS) of nuclear matter at finite temperatures can be also
obtained by the uniform Hartree-Fock approximation, which is shown in
Fig.\ \ref{fig:eos}.
The dotted line is obtained by the Maxwell construction.  Below this line,
the system prefers to split into two parts, namely liquid and gas
phases with different densities, rather than the uniform phase, and
the actual EOS are the dashed lines in this coexistence region.

The caloric curves for the infinite nuclear matter can be also
obtained by the Maxwell construction.
\begin{figure}
\begin{center}
\includegraphics[width=8.5cm,clip]{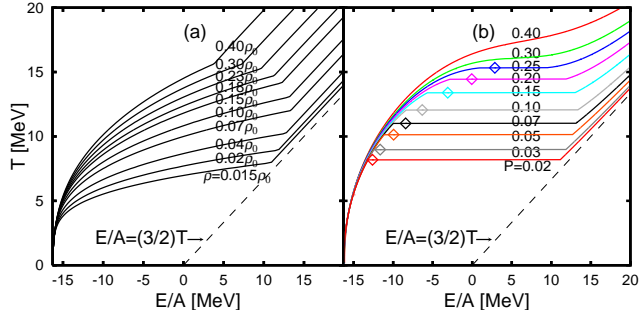}
\end{center}
\caption{The caloric curves of infinite nuclear matter obtained by
  applying the Maxwell construction to the EOS (Fig.\ \ref{fig:eos}).
  The line of $E/A=(3/2)T$ is drawn for the comparison.  (a) The
  constant volume caloric curves drawn for the average densities
  $\rho=0.40\rho_0$, $0.30\rho_0$, $0.23\rho_0$, $0.18\rho_0$,
  $0.15\rho_0$, $0.10\rho_0$, $0.07\rho_0$, $0.04\rho_0$, $0.02\rho_0$
  and $0.015\rho_0$, where $\rho_0=0.17$ fm$^{-3}$ is the normal
  nuclear matter density.  (b) The constant pressure caloric curves
  for the pressures $P=0.02$, 0.03, 0.05, 0.07, 0.10, 0.15, 0.20,
  0.25, 0.30 and $0.40 \,\text{MeV/fm}^3$. The meaning of the diamond
  marks is explained in Sec.\ \ref{CHECKDEC}.}
\label{fig:MFfig}
\end{figure}%
The obtained caloric curves with constant volume and with constant
pressure are shown in Fig.\ \ref{fig:MFfig}(a) and Fig.\
\ref{fig:MFfig}(b), respectively.  The existence of phase transition
can be recognized in both caloric curves.  However, only the constant
pressure caloric curves show temperature plateau at coexistent region.
The constant volume caloric curves show monotonic increase of
temperature with the energy increase even in the coexistent region.
Finding a temperature plateau in the constant pressure caloric curves
is an easy way to identify first order phase transition in infinite
systems.

In the realistic nuclear systems, however, the number of nucleons is
several hundred at most. We have to carefully treat these systems
because the ideas for infinite systems sometimes can not be applied to
such small systems.  Nevertheless, Gross \textit{et al.}
\cite{grossPR,grossPCCP,grossEP} have pointed out that the essence of
phase transition is the anomalous concavity of the entropy $S(E)=\ln
W(E)$, with $W(E)$ being the number of microstates, and the effect of
phase transition can be clearly observed even in finite systems by
finding backbending (negative heat capacity) in the microcanonical
caloric curves.

\section{Framework}
\label{FRAMEWORK}
The microcanonical ensemble with energy $E$, volume $V$ and particle
number $A=N+Z$ can be obtained by solving the time evolution of the
many-nucleon system confined in a container and regarding the state at
each time as a sample of the ensemble.  For the time evolution
calculation, we utilize AMD, which is a microscopic dynamical model
based on the degrees of freedom of interacting nucleons
\cite{onoPRL,onoPTP,onoReview}.  AMD uses a fully antisymmetrized
product of Gaussian wave packets,
\begin{equation}
\langle\mathbf{r}_1\ldots\mathbf{r}_A|\Phi(Z)\rangle=
\det_{ij}\Bigl[\exp\Bigl\{
  -\nu\Bigl(\mathbf{r}_j - \frac{\mathbf{Z}_i}{\sqrt\nu}\Bigr)^2
\Bigr\}
\chi_{\alpha_i}(j)\Bigr],
\label{eq:AMDWF}
\end{equation}
where the complex variables $Z=\{\mathbf{Z}_i;\ i=1,\ldots,A\}$
represent the centroids of the nucleon wave packets.  The width
parameter $\nu$ is treated as a constant parameter and taken as
$\nu=0.16\;\text{fm}^{-2}$ so as to reasonably describe the ground
state of light nuclei such as ${}^{16}\text{O}$.  The spin-isospin
states ($\chi_{\alpha_i}=p\uparrow$, $p\downarrow$, $n\uparrow$, and
$n\downarrow$) are also independent of time. Because of the
antisymmetrization, an AMD wave function $|\Phi(Z)\rangle$ contains
many quantum features so that it is even utilized for the study of
nuclear structures \cite{enyo}.

The time evolution of the centroids $Z$, which parameterize
$|\Phi(Z)\rangle$, is determined by a stochastic equation of motion
\begin{equation}
\frac{d}{dt}\mathbf{Z}_i=\{\mathbf{Z}_i,\mathcal{H}\}
+\Delta\mathbf{Z}_i,
\label{eq:EOMsim}
\end{equation}
where $\{\mathbf{Z}_i,\mathcal{H}\}$ is the deterministic term derived
from the time-dependent variational principle and $\Delta\mathbf{Z}_i$
is the stochastic term, which is important to describe the appearance
of various reaction channels in multifragmentation, for example. The
stochastic term $\Delta\mathbf{Z}_i$ is also essential for the
consistency with the quantum statistics
\cite{ohnishiPRL,ohnishiAP,onoAMDMF,onoSts}.  Two origins of the
stochastic term $\Delta\mathbf{Z}_i$ are considered; one is the
two-nucleon collisions and the other is related to the unrestricted
single-particle motion in the mean field and the localization of
single-particle wave functions when fragments are formed
\cite{onoAu,onoWithFrance,onoReview}.

In the following subsections, we briefly explain these terms referring
to Refs.\ \cite{onoAu,onoWithFrance,onoReview}.  In this paper, we
introduce several new improvements which are necessary for the present
application because the energy conservation and the motion of emitted
nucleons should be treated very consistently.

\subsection{AMD time evolution}
\subsubsection{Deterministic part of the equation of motion}
\label{DETERMINISTIC}
The deterministic term is derived from the time-dependent variational
principle and given by
\begin{equation}
\{Z_{i\sigma},\mathcal{H}\}
=\frac{1}{i\hbar}\sum_{j\tau}C^{-1}_{i\sigma,j\tau}
\frac{\partial\mathcal{H}}{\partial Z^\ast_{j\tau}},
\label{eq:poisson}
\end{equation}
where 
\begin{equation}
C_{i\sigma,j\tau}=\frac{\partial^2} {\partial Z^\ast_{i\sigma}\partial
Z_{j\tau}} \ln\langle\Phi(Z)|\Phi(Z)\rangle
\end{equation}
with $\sigma,\tau=x,y,z$ \cite{onoPRL,onoPTP,onoReview}.  The derived
equation contains the expectation value of the effective Hamiltonian
$\mathcal{H}$, which is given by

\begin{equation}
\mathcal{H}(Z)=\frac{\langle\Phi(Z)|H|\Phi(Z)\rangle}
{\langle\Phi(Z)|\Phi(Z)\rangle}
+\mathcal{K}_{\text{mod}}(Z)+\mathcal{V}_{\text{mod}}(Z).
\label{eq:effectiveH}
\end{equation}
Equation (\ref{eq:effectiveH}) contains the modification terms
$\mathcal{K}_{\text{mod}}(Z)$ and $\mathcal{V}_{\text{mod}}(Z)$ for
the kinetic and potential energies, respectively, both of which are
for the treatment of the gaseous nucleons.  If a nucleon is located
out of nuclei and there are almost no other nucleons around it, the
wave function of such a nucleon is interpreted to have a sharp
momentum distribution rather than the momentum distribution
corresponding to the Gaussian wave packet in Eq.\ \eqref{eq:AMDWF}.
This change of interpretation has been necessary for the consistency
of Q-values of nucleon emission and fragmentation
\cite{onoPTP,onoPRL,onoReview}, and is very important for the
definition of temperature in the present work.  By this modification,
the zero-point kinetic energies of gaseous nucleons are subtracted
from the total kinetic energy \cite{onoPTP,onoPRL,onoReview} by
\begin{equation}
\mathcal{K}_{\text{mod}}(Z)=-\frac{3\hbar^2\nu}{2M}A
+T_0(A-\mathcal{N}_F(Z)),
\label{eq:kmod}
\end{equation}
where the zero-point kinetic energies of isolated fragments are also
subtracted.  The function $\mathcal{N}_F(Z)$ stands for the number of
isolated fragments and nucleons which are assumed to have the definite
center-of-mass momenta without zero-point energies.  Therefore we take
$T_0(A-\mathcal{N}_F)$ as the physical zero-point kinetic energy
instead of $(3\hbar^2\nu/2M)A$.  The functional form of
$\mathcal{N}_F(Z)$ is given in Appendix \ref{ZEROPOINT}.  The
parameter $T_0$ should be $3\hbar^2\nu/2M$ in principle, but it may be
adjusted to fine-tune the binding energies of nuclei.  With the Gogny
force, the binding energies for a wide range of nuclear chart are
reproduced reasonably well by taking $T_0=9.20$ MeV which is close to
$3\hbar^2\nu/2M=10.0$ MeV \cite{onoC12,onoReview}. In Appendix\
\ref{ZEROPOINT}, the degree of isolation $\mathcal{I}_i$ for a nucleon
$i$ is introduced. Using $\mathcal{I}_i$, the total zero-point kinetic
energy can be written as
\begin{equation}
T_0(A-\mathcal{N}_F)=T_0\sum_{i=1}^A(1-\mathcal{I}_i)
\label{eq:zeroForI}
\end{equation}
so that the physical zero-point kinetic energy associated to the
nucleon $i$ may be considered as $T_0(1-\mathcal{I}_i)$.  The
potential energy should be also modified following the same change of
interpretation of gaseous nucleons, because the spatial distribution
of each gaseous nucleon is wide correspondingly to the sharp momentum
distribution.  The details of the modification term
$\mathcal{V}_{\text{mod}}(Z)$ are also given in Appendix
\ref{ZEROPOINT}.  The parameters of the modification terms are chosen
in such a way that a nucleon is treated as a gaseous nucleon (with a
sharp momentum width) if the number of other nucleons around it is
less than about 2.  The binding energies of small nuclei such as
deuteron and triton are utilized to fix some parameters.
 
\par Equation (\ref{eq:effectiveH}) defines the total energy that is
to be conserved by the deterministic term
$\{\mathbf{Z}_i,\mathcal{H}\}$.  The same energy is conserved also by
the stochastic term.  However, we need to be careful in treating the
term $\mathcal{K}_{\text{mod}}(Z)$ because the force
$-T_0\{\mathbf{Z}_i,\mathcal{N}_F\}$ originating from this term is not
always physically reasonable (see Sec.\ \ref{CONSERVATION}).

\subsubsection{Stochastic treatment of single particle motion}
\label{STOCHASTIC}
The stochastic term $\Delta\mathbf{Z}_i$ in Eq.\ (\ref{eq:EOMsim})
consists of two contributions.  The first one is the stochastic
two-nucleon collision term \cite{onoPRL,onoPTP,onoReview} which is not
shown explicitly in the following formulation for simplicity.  The
second one is related to the change of the phase space distribution by
the mean field propagation.  We follow the formalism of AMD/DS given
in Refs.\ \cite{onoWithFrance,onoReview}, but also introduce new
improvements which are crucial in the present study.

We write the Wigner function of the nucleon $k$ as the mean of the
stochastic phase space distributions of deformed Gaussian shape,
\begin{align}
f_k(x,t)&=\overline{g(x; X_k(t), S_k(t))}\\
&=\int{g(x; X, S)w_k(X,t)\frac{d^6X}{\pi^3}}
\label{eq:meanvalue}
\end{align}
with
\begin{multline}
g(x;X,S)=\frac{1}{8\sqrt{\det S}}\\
\times\exp{\Bigl[ -\frac{1}{2}\displaystyle\sum_{a,b=1}^6
      S_{ab}^{-1}(x_a-X_a)(x_b-X_b)\Bigr]}.
\label{eq:deformedWP}
\end{multline}
We have introduced the 6-dimensional phase space coordinates
\begin{equation}
x=\{x_a\}_{a=1,\dots,6}=\left\{\sqrt{\nu}\mathbf{r},
\;\frac{\mathbf{p}}{2\hbar\sqrt{\nu}}\right\}.
\end{equation}
At an initial time, $f_k$ is represented by a single Gaussian wave
packet $g(x;X_k,S_k)$ with $S_{kab}=\frac{1}{4}\delta_{ab}$ and the
wave packet centroid $X_k$ is identified with the physical coordinate
$\mathbf{W}_k$ \cite{onoPRL,onoPTP,onoReview};
\begin{equation}
X=\{X_{ka}\}_{a=1,\dots,6}=
\left\{\text{Re}\mathbf{W}_k,\;\text{Im}\mathbf{W}_k\right\}.
\end{equation}
For a short time, the time evolution of $g_k(x,t)\equiv g(x;X_k(t),
S_k(t))$ by the Vlasov equation
\begin{equation}
\frac{\delta g_k}{\delta t}=-\frac{\partial h}{\partial \mathbf{p}}\cdot
\frac{\partial g_k}{\partial \mathbf{r}}
+\frac{\partial h}{\partial \mathbf{r}}\cdot
\frac{\partial g_k}{\partial \mathbf{p}},
\label{eq:VlasovEACH}
\end{equation}
is characterized by the time evolution of the first and the second
moments of the distribution
\begin{align}
\frac{\delta}{\delta t}X_{ka}(t)
&=\frac{\delta}{\delta t}\int{x_ag_k(x,t)\frac{d^6x}{\pi^3}},\\
\frac{\delta}{\delta t}S_{kab}(t)
&=\frac{\delta}{\delta t}\int{\!\!(x_a-X_{ka}(t))(x_b-X_{kb}(t))g_k(x,t)
\frac{d^6x}{\pi^3}},
\end{align}
in which $(\delta/\delta t)g_k(x,t)$ is given by Eq.\
(\ref{eq:VlasovEACH}). A special notation of the time derivative
$(\delta/\delta t)$ is adopted for the mean field propagation. The
mean distribution [Eq.\ (\ref{eq:meanvalue})] does not change when a
part of the time evolution of the shape $S_{kab}$ is converted into a
stochastic Gaussian fluctuation $\Delta X_{ka}(t)$ to the centroid
$X_{ka}(t)$ as
\begin{align}
\label{eq:EOMX}
\frac{d}{dt}X_{ka}(t)&=\frac{\delta}{\delta t}X_{ka}(t)+\Delta X_{ka}(t)\\
\label{eq:meanDX}
&\overline{\Delta X_{ka}(t)}=0,\\
&\overline{\Delta X_{ka}(t)\Delta X_{kb}(t^\prime)}
=D_{kab}(t)\delta(t-t^\prime),
\label{eq:widthDX}
\end{align}
where $D_{kab}(t)$ denotes the strength and correlation of
fluctuations.  Correspondingly, the equation of motion for
$S_{kab}(t)$ is given by
\begin{align}
\frac{d}{dt}S_{kab}(t)&=\frac{\delta}{\delta t}S_{kab}(t)
-D_{kab}(t).
\label{eq:EOMS}
\end{align}
In order that the Vlasov equation is satisfied, the choice of
$D_{kab}(t)$ is arbitrary as far as the positive definiteness of
$D_{kab}(t)$ and $S_{kab}(t)$ is guaranteed.

In the original version of AMD/DS in
Refs.\ \cite{onoWithFrance,onoReview}, $D_{kab}$ was chosen as the
component diffusing beyond the original width of the wave packet,
that is,
\begin{equation}
D^{(1)}_{kab}(t)
=\lim_{\Delta t \rightarrow 0}\frac{1}{\Delta t}\sum_c\max
\left(0,\;\lambda_c-\frac{1}{4}\right)O_{ac}O_{bc},
\label{eq:diffusionS}
\end{equation}
where $O_{ab}$ and $\lambda_a$ are the diagonalizing orthogonal matrix
and the eigenvalues of the symmetric matrix
$S_{k}(t)+\frac{\delta}{\delta t}S_{k}\Delta t$.  With this original
choice, however, there is no accordance between the momentum spreading
of the distribution $S_{k }$ and the zero-point kinetic energy
$T_0(1-\mathcal{I}_k)$ assumed in the conserved energy (Sec.\
\ref{DETERMINISTIC}).  This accordance is important for the full
consistency of the energy conservation and the precise evaluation of
the temperature in the present work. Therefore we utilize the
arbitrariness of $D$ and introduce another choice $D=D^{(1)}+D^{(2)}$
by adding a term $D^{(2)}$ as explained below.

Ignoring the effect of antisymmetrization for simplicity, the
expectation value of the kinetic energy of an emitted nucleon $k$
can be written as
\begin{equation}
\mathcal{K}_\text{hamil}=\frac{2\hbar^2\nu}{M}
(X_{k4}^2+X_{k5}^2+X_{k6}^2) +T_0(1-\mathcal{I}_k),
\label{eq:khamil}
\end{equation}
which is a part of the conserved energy $\mathcal{H}$ and includes the
zero-point kinetic energy.  On the other hand, if we use the phase
space distribution $g(x;X_k,S_k)$, the kinetic energy is given by
\begin{equation}
\mathcal{K}_\text{distr}
=\frac{2\hbar^2\nu}{M}
(X_{k4}^2+X_{k5}^2+X_{k6}^2)+
\frac{4}{3}T_0\mathop{\mathrm{Tr}_p}S_k,
\label{eq:kdistr}
\end{equation}
where ($3\hbar^2\nu/2M$) has been replaced by $T_0$ consistently with
Eq.\ (\ref{eq:kmod}) and a special notation $\mathop{\mathrm{Tr}}_p
S_k\equiv S_{k44}+S_{k55}+S_{k66}$ has been used.  If only $D^{(1)}$
is taken, $(1-\mathcal{I}_k)$ usually decreases faster than
$\frac{4}{3}\mathop{\mathrm{Tr}}_p S_k$, when the nucleon $k$ is going
out of a nucleus.  Therefore, in order to keep the accordance of
$\mathcal{K}_\text{hamil}$ and $\mathcal{K}_\text{distr}$, we choose
$D_k$ so as to satisfy the condition
\begin{equation}
\mathop{\mathrm{Tr}_p} S_k=\frac{3}{4}
(1-\mathcal{I}_k),
\label{eq:CONWIDTH}
\end{equation}
when the right hand side is getting smaller than the left hand side.
This requirement can be satisfied by scaling $S_{kab}$ as
\begin{equation}
S_{kab}(t+\Delta t)=R_kS'_{kab}
\end{equation}
by a factor
\begin{equation}
R_k=\text{min}\left\{\frac{
\frac{3}{4}(1-\mathcal{I}_k)}
{\mathop{\mathrm{Tr}_p}S^\prime_k},\;1\right\},
\label{eq:CONWIDTHfactor}
\end{equation}
where $S^\prime_{kab}=S_{kab}(t)+(\frac{\delta}{\delta
  t}S_{kab}-D^{(1)}_{kab})\Delta t$.  This corresponds to taking the
choice $D_{kab}(t)=D^{(1)}_{kab}(t)+D^{(2)}_{kab}(t)$ with
\begin{equation}
D^{(2)}_{kab}(t)=\lim_{\Delta t \rightarrow 0}\frac{1}{\Delta t}
(1-R_k)S'_{kab}.
\end{equation}

\subsubsection{Decoherence process}
\label{DECOHERENCE}
The fluctuation $\Delta X$ has been introduced in Eq.\ (\ref{eq:EOMX})
by the condition that the single-particle dynamics of the mean field
propagation would be reproduced. However, the mean field propagation
is not sufficient to approximate the time evolution of many-body
systems, at least, in the sense that the idempotency of the one-body
density matrix ($\hat{\rho}^2=\hat{\rho}$) is unphysically kept during
the mean field propagation. After enough time has past in the
multifragmentation reactions, for example, the reduced one-body
density matrix would be rather represented by an ensemble of density
matrices in each of which single-particle wave functions are localized
in fragments.  This means that the coherence of the single-particle
state is lost at some time (quantum branching).  The decoherence is
due to the many-body correlations and therefore its time scale
(coherence time $\tau_0$) should be related to many-body effects in
some way.

In Refs.\ \cite{onoWithFrance,onoReview}, the decoherence is
assumed to take place for a nucleon when it is scattered by a
two-nucleon collision, which is the effect beyond mean field.  In the
present calculation, however, we choose a different prescription to
investigate the dependence on the decoherence process.

For each nucleon $k$, if there are more than three other nucleons
within the radius of 2 fm (measured in
$\mathop{\mathrm{Re}}\mathbf{W}/\sqrt{\nu}$), a decoherence process is
assumed to take place with the probability of $1/\tau_0$ per unit
time.  When a decoherence process takes place, it affects all the
nucleons $i$ that are located within the radius of 2 fm including the
nucleon $k$ itself and each of the single particle wave packet of the
nucleon $i$ is replaced by a Gaussian wave packet with the phase space
distributions
\begin{equation}
S_{iab}=
\begin{cases}
\frac{1}{4} & (a=b=1,2,3)\\
\frac{1}{4}(1-\mathcal{I}_i) & (a=b=4,5,6)\\
0 & (a\neq b)
\end{cases}.
\end{equation}
The momentum widths are replaced by $\frac{1}{4}(1-\mathcal{I}_i)$
rather than the standard Gaussian width $\frac{1}{4}$ in order to
satisfy Eq.\ (\ref{eq:CONWIDTH}).  It should be noted that a single
decoherence process affects several nucleons at once and therefore the
rate of decoherence for a specific nucleon is approximately
proportional to the number of the neighboring nucleons.  We take
$\tau_0=500$ fm/$c$ for usual calculations in the present work, but we
also check the dependence of the results on $\tau_0$ in Sec.\
\ref{CHECKDEC}.

\subsubsection{The stochastic equation of motion}
\label{CONSERVATION}

The stochastic equation of motion for the wave packet centroids $Z$ is
given by \cite{onoAu,onoReview}
\begin{align}
\frac{d}{dt}\mathbf{Z}_i&=\{\mathbf{Z}_i,\mathcal{H}\}
+\sum_{k=1}^A\Bigr((\Delta\mathbf{Z}_i)^{(k)}_\text{flct}
+(\Delta\mathbf{Z}_i)^{(k)}_\text{dssp}\Bigr),
\label{eq:EOM}\\
&(\Delta\mathbf{Z}_i)^{(k)}_\text{flct}
=\left\{\mathbf{Z}_i,\;\mathcal{O}^\prime_k
+T_0\mathcal{I}_k\right\}_{\mathrm{C}_k},
\label{eq:EOMfluc}\\
&(\Delta\mathbf{Z}_i)^{(k)}_\text{dssp}
=\mu_k\left(\mathbf{Z}_i,\mathcal{H}^\prime
\right)_{\mathrm{N}_k}.
\label{eq:EOMdssp}
\end{align}

The fluctuation term $(\Delta\mathbf{Z}_i)^{(k)}_\text{flct}$ is
obtained by converting the fluctuation $\Delta X_{ka}$ of the physical
coordinate $W$ to that of the original AMD coordinates $Z$.  This
conversion is done \cite{onoReview,onoAu} by introducing an
expectation value $\mathcal{O}_k(Z,t)$ of a stochastic one-body
operator which generates the fluctuation as
$\{\mathbf{Z}_i,\mathcal{O}_k\}$ by the Poisson brackets [Eq.\
(\ref{eq:poisson})].  $\mathcal{O}^\prime_k$ in Eq.\
(\ref{eq:EOMdssp}) is different from $\mathcal{O}_k$ by the Lagrange
multiplier terms for the conservation of the three components of the
center-of-mass coordinate and those of the total momentum
\cite{onoReview,onoAu}.  The term $T_0\{\mathbf{Z}_i,\mathcal{I}_k\}$
in $(\Delta\mathbf{Z}_i)^{(k)}_\text{flct}$ is introduced in order to
cancel the unphysical force $-T_0\{\mathbf{Z}_i,\mathcal{I}_k\}$ in
the deterministic term $\{\mathbf{Z}_i,\mathcal{H}\}$. Such a force
does not exist for the single-particle motion in the mean field
one-body dynamics.  The subtraction is introduced as a term in
$(\Delta\mathbf{Z}_i)^{(k)}_\text{flct}$ for a technical reason
(related to the dssp term) so that the Hamiltonian $\mathcal{H}$ with
the zero-point correction term $T_0(1-\mathcal{I}_k)$ is still the
conserved energy.

The term $(\Delta\mathbf{Z}_i)^{(k)}_\text{dssp}$ is the dissipation
term to achieve the energy conservation, where
$(Z_{i\sigma},\mathcal{H}^\prime)$ is defined by
\begin{equation}
(Z_{i\sigma},\mathcal{H}^\prime)
=\frac{1}{\hbar}\sum_{j\tau}C^{-1}_{i\sigma,j\tau}
\frac{\partial\mathcal{H}^\prime}{\partial Z^\ast_{j\tau}}.
\end{equation}
The coefficient $\mu_k$ for each $k$ is determined so that the energy
violation by the $(\Delta\mathbf{Z}_i)^{(k)}_\text{flct}$ term is
compensated by $(\Delta\mathbf{Z}_i)^{(k)}_\text{dssp}$.  Lagrange
multiplier terms are included in the effective Hamiltonian
$\mathcal{H}'$, assuming that $(\Delta\mathbf{Z}_i)^{(k)}_\text{dssp}$
conserves some global one-body quantities such as the center-of-mass
coordinate, the total momentum, the total orbital angular momentum,
and the monopole and quadrupole moments in the coordinate and momentum
spaces \cite{onoAu,onoReview}. 

The subscripts $\mathrm{C}_k$ and $\mathrm{N}_k$ mean that the
contents of the brackets are calculated by limiting to the subsystem
$\mathrm{C}_k$ or $\mathrm{N}_k$. $\mathrm{C}_k$ is the cluster that
includes the nucleon $k$, where the clusters are identified by the
condition that two nucleons $i$ and $j$ belong to the same cluster if
$|\mathbf{Z}_i-\mathbf{Z}_j|<1.75$. $\mathrm{N}_k$ stands for a
neighborhood of nucleon $k$ defined by
\begin{align}
\mathrm{N}_k&=\{i;\;|\mathbf{W}_i-\mathbf{W}_k|<2.5,
i\in \mathrm{C}_k,
i\neq k,
\ \mbox{and}\ M_i>1 \},
\label{eq:Nk}
\end{align}
where $M_i$ is the number of nucleons within the distance of 3 fm
(measured in $\mathop{\mathrm{Re}}\mathbf{W}/\sqrt{\nu}$) from nucleon
$i$. These limitations ensure that the fluctuation $X_{ka}$ for the
nucleon $k$ should affect only the nucleons within the interactive
range of the nucleon $k$ when the conservation laws are imposed.  The
condition $M_i>1$ in Eq.\ (\ref{eq:Nk}) is introduced to exclude
gaseous nucleons (here we regard the nucleon $i$ with $M_i\leq 1$ as
gaseous) from $\mathrm{N}_k$ so that the dynamics of the gaseous
nucleon $i$ faithfully follows the single-particle motion determined
by the deterministic term $\{\mathbf{Z}_i,\mathcal{H}\}$ and the
fluctuation term
$\sum_{k=1}^A(\Delta\mathbf{Z}_i)_{\text{flct}}^{(k)}$. The choice of
$\mathrm{C}_k$ and $\mathrm{N}_k$ here is similar to Refs.\
\cite{onoAu,onoReview}, but has been updated in order to
carefully treat gaseous nucleons, which is required for the precise
definition of temperature.

We cancel the fluctuation for a nucleon $k$ when the number of
nucleons in $\mathrm{N}_k$ is less than five to prevent small clusters
from unphysically breaking.

\subsection{Reflection at the wall}
\label{REFLECTION}
To obtain a microcanonical ensemble with a fixed volume, we need to
keep nucleons inside the container with a given volume
$V=\frac{4\pi}{3}r^3_\text{wall}$ during the time evolution.  We
introduce a kind of reflection process for this purpose, when nucleons
or fragments are going out of the container.  At each time step after
the time evolution without any effect of the container wall, we judge
whether each nucleon $k$ is in the container
($\frac{1}{\sqrt{\nu}}|\text{Re}\mathbf{W}_k|\leq r_\text{wall}$).  If
an isolated nucleon $k$ is located outside the container and its
momentum $\mathbf{P}_k=2\hbar\sqrt{\nu}\text{Im}\mathbf{Z}_k$ directs
outward, we change the momentum direction
$\hat{\mathbf{P}}_k=\mathbf{P}_k/|\mathbf{P}_k|$ into an inward
direction $\hat{\mathbf{P}}^\prime_k$ which satisfies
$\mathbf{R}_k\cdot\hat{\mathbf{P}}^\prime_k<0$, where
$\mathbf{R}_k=\frac{1}{\sqrt{\nu}}\mathrm{Re}\mathbf{Z}_k$. We
randomly choose the direction $\hat{\mathbf{P}}^\prime_k$ as in the
case of the reflection by an irregular surface. The absolute value of
the momentum $|\mathbf P^\prime_k|$ is adjusted so as to conserve the
total energy which is sometimes affected by antisymmetrization. The
total angular momentum of the system is not conserved because of the
irregular reflection, which allows us to construct a microcanonical
ensemble without the constraint of the total angular momentum.

When a nucleon which belongs to a cluster, where we regard the nucleon
$i$ and $j$ belong to the same cluster when
$|\mathbf{W}_i-\mathbf{W}_j|<0.8$, is located outside the container,
we apply a similar reflection procedure to the center of mass
coordinate of the cluster.  If one applied the reflection procedure to
each of the nucleons in the cluster as is done in
Refs. \cite{ohnishiPRL,ohnishiAP,FMDLGpt,onoAMDMF,onoSts,sugawaVolkov,sugawaGogny},
the cluster would be unphysically broken by the crash to the wall.

When a nucleon $k$ is reflected, the shape $S_k$ has to be
reflected consistently with the change of the momentum direction
$\hat{\mathbf{P}}_k\to\hat{\mathbf{P}}^\prime_k$.  We consider the
reflection with respect to the plane that includes the point
$\mathbf{R}_k$ and is perpendicular to
$\hat{\mathbf{n}}_k=(\hat{\mathbf{P}}^\prime_k-\hat{\mathbf{P}}_k)/
|\hat{\mathbf{P}}^\prime_k-\hat{\mathbf{P}}_k|$, and hence the component
parallel to $\hat{\mathbf{n}}_k$ is reversed. According to this
operation, the coordinate or momentum vector $\mathbf{x}_k$ in the
intrinsic frame of the wave packet $k$ is transformed into
\begin{gather}
\mathbf{x}^\prime_k=\mathbf{x}_k-2\left(\mathbf{x}_k\cdot
\hat{\mathbf{n}}_k\right)\hat{\mathbf{n}}_k
=T_k\;\mathbf{x}_k,\\ 
(T_k)_{\sigma\tau}
=\delta_{\sigma\tau}-2n_{k\sigma}n_{k\tau} \quad(\sigma,\tau=x,y,z).
\end{gather}
By defining the transformation matrix in the phase space as
\begin{equation}
\mathcal{T}_k
=\begin{pmatrix}
T_k & 0\\
0 & T_k
\end{pmatrix},
\end{equation}
the transformation of the shape $S_k$ is given by
\begin{equation}
S'_{kab}=\sum_{cd}(\mathcal{T}_k)_{ac}(\mathcal{T}_k)_{bd} S_{kcd}.
\end{equation}

In the actual calculation, for the purpose of numerical stability, we
introduce a small delay time $\tau_{\text{delay}}$ for the response to
the fluctuation $\Delta X_{ka}$ \cite{onoAu,onoReview}.  This
delayed fluctuation $\Xi_{ka}(t)$ also has to be transformed as
$\Xi^\prime_{ka}=\sum_b(\mathcal{T}_k)_{ab}{\Xi}_{kb}$ when a nucleon
$k$ is reflected.

\section{Calculation of temperature and pressure}
\label{CALC}
In order to obtain caloric curves, we need to extract the temperature
$T$ of each microcanonical ensemble $\{E,V,A\}$. We also need to know
the pressure $P$ of the ensemble in order to draw the constant
pressure caloric curves. In this section, we explain how to obtain $T$
and $P$ in our calculation.

The microcanonical temperature is defined by $T^{-1}=(\partial
S/\partial E)|_{V,A}$. This quantity can be evaluated with a good
precision by using the average kinetic energy of gaseous nucleons.
When the total volume is $V>A/\rho_0$, where $\rho_0=0.17$
fm$^{-3}$ is the normal nuclear density, it is possible to find a
partial system composed of gaseous nucleons to which classical ideal
gas relations are applicable. Let us call this partial system
$\mathrm{G}$ which will be defined below.  Appendix\ \ref{ISOTEMP}
shows that the microcanonical temperature $T$ of the ensemble
$\{E,V,A\}$ can be obtained from the average kinetic energy of
G-subsystem as
\begin{equation}
T=\frac{2}{3}\biggl<
\frac{\mathcal{K}_\mathrm{G}}{A_\mathrm{G}}
\biggr>_{\{E,V,A,A_\mathrm{G}>0\}},
\label{eq:microTapp}
\end{equation}
where $\mathcal{K}_\mathrm{G}$ and $A_\mathrm{G}$ are the kinetic
energy and the number of nucleons of G-subsystem, respectively,
and the brackets $\langle\,\rangle_{\{E,V,A,A_\mathrm{G}>0\}}$ denote
the average value for the microcanonical ensemble with
$A_\mathrm{G}>0$.  Because of the consistency of Eq.\
(\ref{eq:khamil}) and Eq.\ (\ref{eq:kdistr}), $\mathcal{K}_\mathrm{G}$
can be calculated as
\begin{equation}
\mathcal{K}_\mathrm{G}=\sum_{k\in\mathrm{G}}\Bigl\{
\frac{2\hbar^2\nu}{M}
(X_{k4}^2+X_{k5}^2+X_{k6}^2)
+\frac{4}{3}T_0\mathop{\mathrm{Tr}_p}S_k\Bigr\}.
\label{eq:kinE}
\end{equation}
The first term is the contribution from the wave packet centroid. The
accumulated delayed fluctuation $\tau_{\text{delay}}\mathbf{\Xi}_k$
\cite{onoAu} should be included in this term because the corresponding
shape shrinking has been already applied. The second term is the
contribution from the momentum widths of $S_k$.

According to Appendix\ \ref{ISOTEMP}, G-subsystem can be defined
arbitrarily as far as the following two conditions are satisfied. The
first condition is that the quantum effect should be negligible for
any nucleon in G-subsystem. The second condition is that the
G-subsystem should be selected without using momentum variables, i.e.,
the configurations with different nucleon momenta should be equally
taken into account if they have the same nucleon positions.

In our actual calculation, G-subsystem has been chosen in the
following way. We select the nucleons $k$ for which the density of the
nucleons with the same spin-isospin
\begin{equation}
\rho_\alpha=\sum_{i\in\alpha_k ,i\neq
k}\left(\frac{2\nu}{\pi}\right)^{3/2}
e^{-2(\text{Re}\mathbf{W}_k-\text{Re}\mathbf{W}_i)^2},
\end{equation}
is sufficiently low ($\rho_\alpha\leq\rho_\mathrm{G}=(1/200)\rho_0$)
so that the antisymmetrization effect can be neglected. It is
necessary to eliminate the nucleons belonging to clusters because the
discrete level(s) of the internal degrees of freedom can only be
treated quantum mechanically.  Therefore, among the selected nucleons,
we choose the nucleons $k$ which do not have more than one other
nucleon within the distance of $r_\mathrm{G}=3\,\text{fm}$ (if we use
$M_k$ defined in Sec.\ \ref{STOCHASTIC}, we choose the nucleon $k$
which satisfy the condition $M_k\leq 1$).  Because of our reflection
procedure, it is possible that nucleons locate outside the container
in short time interval. Those nucleons are excluded from the selected
nucleons. The above selections are done based on spatial coordinates
without any momentum selection.  We define the system composed of
these selected nucleons as G-subsystem.  The results should be
independent of the definition of G as far as the necessary conditions
are respected, which will be checked in Sec.\ \ref{CHECKDEC} by
changing the criteria in our definition of G.

For the calculation of the pressure $P$, we adopt the common
definition of pressure as the external force necessary to keep the
volume. Namely the pressure is given by
\begin{equation}
P=\frac{2}{4\pi r^2_{\text{wall}}\tau_{\text{total}}}
\sum_\text{reflections}\Delta\mathbf{p}\cdot\hat{\mathbf{N}}
\end{equation}
where the summation is taken over all the reflections at the container
wall (see Sec.\ \ref{REFLECTION}) which occurred during the time
evolution of the total time $\tau_{\text{total}}$, $\Delta\mathbf{p}$
is the momentum change at each reflection, and $\hat{\mathbf{N}}$ is
the normal vector.  Factor two is from the fact that when a nucleon or
a fragment hits the wall the total momentum of the rest of the system
is also changed for the momentum conservation.

\section{The studied system and obtained caloric curves}
\label{RESULT}
\subsection{The studied system}
\label{SYSTEM}
In the present study, the system with $(N,Z)=(18,18)$ is considered,
which is the same system taken in Refs.
\cite{sugawaVolkov,sugawaGogny}.  There are several examples
indicating that phase transition exists even in such small systems
\cite{schmidt, gobet1,gobet2}.

Microcanonical ensembles are constructed with the radius of the
container $r_{\text{wall}}=5$, $5.5,\dots,15$ fm, which correspond to
the average densities $A/(\frac{4\pi}{3}r_\text{wall}^3)=0.40\rho_0$,
$0.30\rho_0,\dots,0.015\rho_0$, and the energy $E^\ast/A=4,$ $\
6,\dots,28$ MeV, where $E^\ast$ stands for the excitation energy
relative to the ground state of ${}^{36}\text{Ar}$ nucleus
($E_{\text{g.s.}}=-8.9A$ MeV).  The minimum radius $r_{\text{wall}}=5$
fm corresponds to a volume of the container $V=2.5A/\rho_0$, and
therefore we do not consider a compressed liquid nucleus in this
study.

The time evolution was calculated up to 20000 fm/$c$, which is much
longer than a typical nuclear reaction time scale ($\sim 100$
fm/$c$). The states for the first 5000 fm/$c$ were discarded in order
to remove the initial state dependence. Samplings were done at every
10 fm/$c$. Similar calculations were carried out 4 times independently
to improve the statistics.

\subsection{Obtained caloric curves}
\label{CALORIC}
\begin{figure}
\begin{center}
\includegraphics[width=6cm,clip]{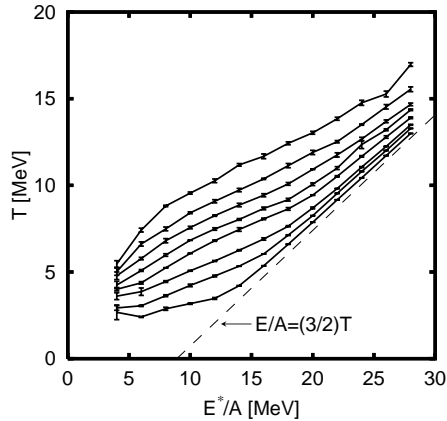}
\end{center}
\caption{The constant volume caloric curves for the $A=36$ system
  obtained by AMD.  The lines correspond to the container size
  $r_\text{wall}=5$, 5.5, 6, 6.5, 7, 8, 9 and 11 fm from the
  top. Statistical uncertainty is shown by error bars. The line of
  $E/A=(3/2)T$ is drawn for comparison.}
\label{fig:vconst}
\end{figure}

The obtained constant volume caloric curves with $r_\text{wall}=5$,
5.5, 6, 6.5, 7, 8, 9 and 11 fm are shown in Fig.\ \ref{fig:vconst}.
The caloric curves with $r_\text{wall}=13$ and 15 fm are also
constructed, but they are not shown in Fig\ \ref{fig:vconst} because
the complete equilibration could not be achieved by our investigation
time (20000 fm/$c$) in the low energy cases ($E^\ast/A=4\sim8$ MeV)
[the results systematically differ if we start the calculation with a
quite different initial state].  Otherwise, fairly smooth caloric
curves are obtained.  The negative heat capacity (backbending) or
plateau does not appear in the constant volume caloric curves, which
is consistent with the result of Ref.\ \cite{sugawaGogny} with another
version of AMD.  It was expected from the fact that the constant
volume caloric curves for infinite nuclear matter do not show the
plateau even in coexistent region (Fig.\ \ref{fig:MFfig}(a)).
 
Although this result (Fig.\ \ref{fig:vconst}) is somehow similar to
the caloric curves for infinite matter (Fig.\ \ref{fig:MFfig}(a)), the
transition from the liquid-gas phase coexistence to the pure gas phase
is not clear. It is difficult to judge clearly whether the phase
transition exist or not from the constant volume caloric curves.  On
the other hand, the constant pressure caloric curves which showed
plateau in the case of infinite nuclear matter (Fig.\
\ref{fig:MFfig}(b)) are expected to show the signal of the liquid-gas
phase transition most clearly.  Therefore, the judgment should be done
with the constant pressure caloric curve, not with the constant volume
caloric curve.

\begin{figure}
\begin{center}
\includegraphics[width=6cm,clip]{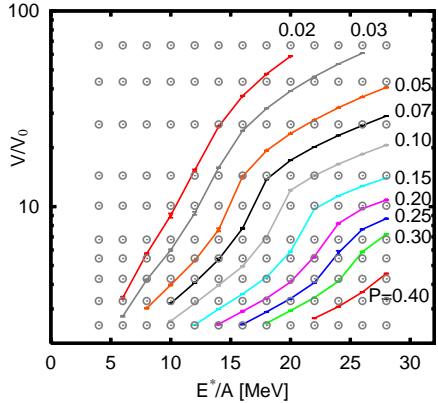}
\end{center}
\caption{The constant pressure lines drawn on $E$-$V$ plain. The
  circles indicate the points where the microcanonical ensembles are
  constructed. The lines correspond to the constant pressure lines
  with $P=0.02$, 0.03, 0.05, 0.07, 0.10, 0.15, 0.20, 0.25, 0.30 and
  0.40 MeV/fm$^3$, which are obtained by performing the transformation
  between volume $V$ and pressure $P$ with the interpolation at each
  energy $E$. Statistical uncertainty is shown by error bars.}
\label{fig:EV}
\end{figure}

Constant pressure caloric curves can also be constructed from our
results as follows. After the evaluation of all the ensembles
$\{E,V,A\}$, we eventually know the temperature and the pressure at
each lattice point on the $E$-$V$ plain indicated by circles in Fig.\
\ref{fig:EV}. From these results, we can draw constant pressure lines
on the $E$-$V$ plain by performing the transformation between volume
$V$ and pressure $P$ with the interpolation at each energy $E$. The
obtained constant pressure lines are shown in Fig.\ \ref{fig:EV}. The
relation between energy $E$ and temperature $T$ along these constant
pressure lines corresponds to the constant pressure caloric curves.
These caloric curves should not be confused with the caloric curves
for the constant pressure ensemble $\{E,P,A\}$. The obtained
caloric curves are those calculated from the microcanonical ensemble
$\{E,V_P,A\}$, where the volume $V_P$ is chosen so that the ensemble
gives a certain pressure $P$.
\begin{figure}
\begin{center}
\includegraphics[width=6cm,clip]{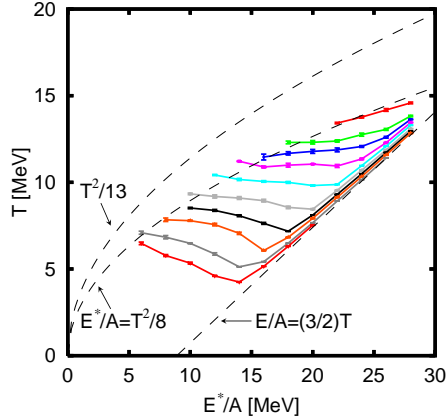}
\end{center}
\caption{The constant pressure caloric curves for the $A=36$ system
  obtained by AMD. The lines correspond to the pressure $P=0.02$, 0.03,
  0.05, 0.07, 0.10, 0.15, 0.20, 0.25, 0.30 and 0.40 MeV/fm$^3$ from the
  bottom. Statistical uncertainty is shown by error bars. The curves
  of $E^\ast/A=T^2/(8\,\text{MeV})$ and
  $E^\ast/A=T^2/(13\,\text{MeV})$, and the line of $E/A=(3/2)T$ are
  drawn for comparison.}
\label{fig:pconst}
\end{figure}%
Figure \ref{fig:pconst} shows the constant pressure caloric curves
with the pressure $P=0.02$, 0.03, 0.05, 0.10, 0.15, 0.20, 0.25, 0.30
and $0.40\;\text{MeV}/\text{fm}^3$ obtained by the above procedure. In
each caloric curve with $P\lesssim 0.15$ MeV/$\text{fm}^3$, negative
heat capacity is observed clearly, which is the signal of first order
phase transition.  For example at $P=0.05$ MeV/$\text{fm}^3$, the
caloric curve start from $E^\ast/A=8$ MeV and the temperature
decreases till the energy of 16 MeV even the system is heated up from
8 MeV to 16 MeV, and after the energy of 16 MeV, the temperature goes
up with a slope 3/2. By the comparison with the infinite nuclear
matter caloric curves (Fig.\ \ref{fig:MFfig}(b)), the caloric curves
with $P<0.20$ MeV/$\text{fm}^3$ in Fig.\ \ref{fig:pconst} can be
interpreted as the caloric curves drawn from the liquid-gas
phase coexistence to the pure gas phase.
\begin{figure}
\begin{center}
\includegraphics[width=8.5cm,clip]{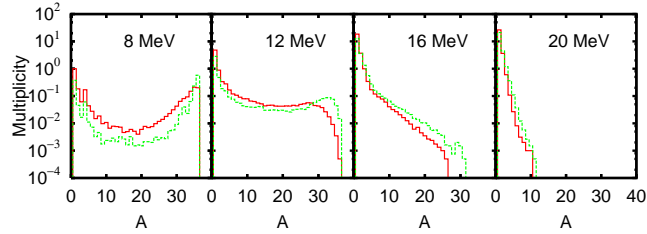}
\end{center}
\caption{The fragment mass distributions along the $P=0.05$ MeV/fm$^3$
  line. Full lines are the distributions obtained with
  $r_\text{clust}=2.5$ fm. Dashed lines are the distributions obtained
  with $ r_\text{clust}=3.0$ fm.}
\label{fig:mass}
\end{figure}
Figure \ref{fig:mass} shows the fragment mass distributions for the
ensembles along the $P=0.05$ MeV/fm$^3$ line. Full lines are the
distributions of the fragments identified by the condition that two
nucleons $i$ and $j$ belong to the same cluster if
$|\frac{1}{\sqrt{\nu}}\text{Re}(\mathbf{W}_i-\mathbf{W}_j)|<
r_\text{clust}=2.5$ fm. These distributions do not necessarily
correspond to the fragments which would be emitted when the container
wall is removed. In fact, the distributions change when
$r_\text{clust}$ is varied. The dashed lines in Fig.\ \ref{fig:mass}
show the fragment mass distributions obtained with
$r_\text{clust}=3.0$ fm.  Nevertheless, these distributions are
helpful for the qualitative understanding of the change of the system
with energy increase.  When the energy is low ($E^\ast/A=8$ MeV), the
distribution shows U-shape with two peaks so that the typical
configuration at this energy is a large nucleus coexisting with a few
gaseous nucleons.  When the energy is increased
($E^\ast/A=12\sim16\,\text{MeV}$), the peak at the large fragment
becomes smaller and the distribution changes into shoulder-like and
power-low-like shapes.  Thus complex configurations with many
intermediate and light mass fragments are typical at these energies,
and the proportion of light fragments increases as the energy
increases (from $E^\ast/A=12\,\text{MeV}$ to 16 MeV).  When the energy
is sufficiently high ($E^\ast/A=20$ MeV), the distribution changes
into exponential shape, which can be interpreted that the nucleons are
moving almost freely at this energy, although a few nucleons come
close and make compounds with small probability. The change of the
distribution is fully consistent with the interpretation of the
caloric curve, that is, the system along the $P=0.05$ MeV/fm$^3$ line
changes from liquid-gas phase coexistence at low energy to pure-gas
phase at high energy.

The caloric curves in Fig.\ \ref{fig:pconst} do not contain the pure
liquid phase because of our choice of the container size
$r_{\text{wall}}\geq 5$ fm. Nevertheless, we can roughly guess the
pure liquid caloric curve from our result as follows. Diamond marks in
the constant pressure caloric curves for infinite nuclear matter
(Fig.\ \ref{fig:MFfig}(b)) indicate the points corresponding to the
density with $r_{\text{wall}}=5$ fm.  These points are inside of the
plateau region, but are very close to the left edge of the coexistence
region.  Furthermore, as is seen in Fig.\ \ref{fig:MFfig}(b), the pure
liquid caloric curves are almost independent of the pressure unless
the liquid is compressed with higher pressure ($P>0.3$
MeV/$\text{fm}^3$). We can guess the corresponding line by connecting
the leftmost points of our AMD result and the line can be compared to
the Fermi gas formula $E^\ast/A=aT^2$ for excited nuclei. The pure
liquid caloric curve should be slightly left to the connected
line. Thus our AMD result seems to be consistent with the quantum and
fermionic statistics with a reasonable level density parameter
$a^{-1}=8\sim13$ MeV.

If we simply define the critical point as the point where the negative
heat capacity disappears, the critical temperature and the critical
pressure can be estimated as $T_\mathrm{C}\sim 12$ MeV and
$P_\mathrm{C}\sim 0.20$ MeV/fm$^3$, respectively. These values are
lower than those of infinite nuclear matter ($T_\mathrm{C}\sim 16$ MeV
and $P_\mathrm{C}\sim 0.3$ MeV/fm$^3$). The lowering of $T_\mathrm{C}$
seems to be reasonable because the existence of the surface for
fragments reduces the interaction energy gain by being
fragments (liquid phase).

\subsection{Check of the theoretical ambiguities}
\label{CHECKDEC}
\begin{figure}
\begin{center}
\includegraphics[width=7.0cm,clip]{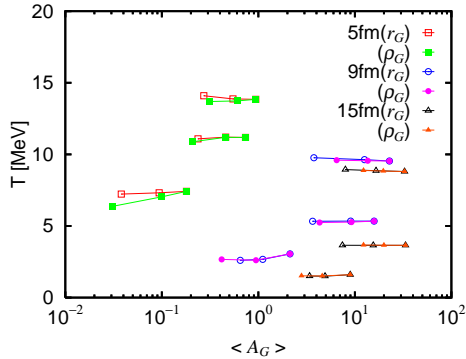}
\end{center}
\caption{The temperatures obtained with different parameter sets
  ($r_\mathrm{G}$ and $\rho_\mathrm{G}$) for choosing G-subsystem. The
  results of the systems with $r_\mathrm{wall}=5,9,15$ fm and
  $E^\ast/A=6,14,22$ MeV are shown. The abscissa $\langle
  A_\mathrm{G}\rangle$ is the average number of nucleons in
  G-subsystem and the parameters are chosen so that $\langle
  A_\mathrm{G}\rangle$ becomes about half and about a quarter of the
  original choice [$r_\mathrm{G}=3$ fm and
  $\rho_\mathrm{G}=(1/200)\rho_0$].  The open points show the
  dependence of $r_\mathrm{G}$ and the solid points show the
  dependence of $\rho_\mathrm{G}$, which are connected by lines to
  guide eyes. Statistical uncertainty is similar to or smaller than the
  size of points.}
\label{fig:checkpara}
\end{figure}

As is mentioned in Sec.\ \ref{CALC}, the measured temperatures should
be independent of the choice of G-subsystem as far as the conditions
described in Sec.\ \ref{CALC} are satisfied. This can be checked by
changing the criteria in the definition of G.  In Sec.\ \ref{CALC},
G-subsystem has been selected as the subsystem of the nucleons which
satisfy $\rho_\alpha\leq\rho_{\mathrm{G}}=(1/200)\rho_0$ and do not
have more than one other nucleon within the distance of
$r_\mathrm{G}=3\,\text{fm}$.

According to Appendix\ \ref{ISOTEMP}, the dependence on
$\rho_\mathrm{G}$ and $r_\mathrm{G}$ should be weak if only exist.
This is actually demonstrated in Fig.\ \ref{fig:checkpara} which shows
the temperatures obtained with different choices of parameters
($r_\mathrm{G}$ and $\rho_\mathrm{G}$) to select G-subsystem.  The
parameters are chosen so that the average number of nucleons in
G-subsystem $\langle A_\mathrm{G}\rangle$ becomes about half and about
a quarter of the original choice.  The open points show the dependence
on $r_\mathrm{G}$ and the solid points show the dependence on
$\rho_\mathrm{G}$.  The temperatures are almost independent of the
choice of the parameters.  Only small dependence on the parameters can
be noticed when $\langle A_\mathrm{G}\rangle\lesssim 1$.  This is
probably because we measure the temperature by using only the states
with $A_\mathrm{G}>0$ (see Appendix\ \ref{ISOTEMP}).  Nevertheless,
the dependence is very weak and seen only at low energy and at small
volume so that the ambiguity associated with this issue would hardly
affect the obtained results shown in Sec.\ \ref{CALORIC} and would not
change the conclusions.

\begin{figure}
\begin{center}
\includegraphics[width=8.5cm,clip]{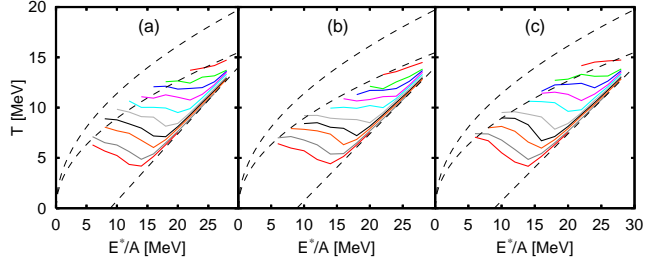}
\end{center}
\caption{The AMD results of the constant pressure caloric curves with
  $P=0.02$, 0.03, 0.05, 0.10, 0.15, 0.20, 0.30 and 0.40
  MeV/$\text{fm}^3$ from the bottom when the coherence time is varied:
  (a) $\tau_0=250$ fm/c, (b) $\tau_0=1000$ fm/c, (c) the same
  procedure as in Refs.\ \cite{onoWithFrance,onoReview}.}
\label{fig:compare}
\end{figure}

In the current AMD dynamical calculation, the coherence time $\tau_0$
(Sec.\ \ref{DECOHERENCE}) is somehow arbitrary. The above results
(Figs.\ \ref{fig:vconst}-\ref{fig:checkpara}) are obtained with the
choice $\tau_0=500\,\text{fm}/c$. We should check the robustness of
the obtained results against the arbitrariness of the coherence time.
Thus we have done the calculations with $\tau_0=250$ and
$1000\,\text{fm}/c$. The obtained constant pressure caloric curves are
shown in Fig \ref{fig:compare}(a) and Fig.\ \ref{fig:compare}(b),
respectively.
Furthermore, we have done the calculation with another choice of the
decoherence process adopted in Refs.\ \cite{onoWithFrance,onoReview}
and the obtained constant pressure caloric curves are shown in Fig.\
\ref{fig:compare}(c).  By the comparison of the figures in Fig.\
\ref{fig:compare} and Fig.\ \ref{fig:pconst}, we notice some change of
the leftmost point of each caloric curve, which can be interpreted as
the change of the level density parameter $a$ for the liquid caloric
curve.  This dependence limits the possible range of the coherence
time $\tau_0$ when the level density parameter $a$ is known.  In any
case, negative heat capacity is observed in all the caloric curves for
a wide range of $\tau_0$.

\section{Summary}
\label{SUMMARY}
In this paper, we have constructed the microcanonical caloric curves
of a system with 36 nucleons [$(N,Z)=(18,18)$] using the microscopic
time evolution of AMD.  We applied the same AMD model that has been
applied to nuclear reactions successfully.

For the first time as a molecular dynamics calculation for nuclear
systems, we have constructed the constant pressure caloric curves and
found that negative heat capacity appears there with an appropriate
nuclear force (the Gogny force).  Negative heat capacity (backbending)
is a specific character of first order phase transition in finite
systems. Thus we confirm the existence of nuclear liquid-gas phase
transition under the ideal thermal equilibrium condition with AMD.
The obtained fragment mass distributions also support the existence of
the phase transition.  We have checked that the extracted temperature
from the ensembles do not change when different criteria to choose the
nucleons for the temperature measurement are used as long as necessary
conditions are satisfied.  We have also checked the obtained
conclusions are insensitive to the theoretical ambiguity, i.e. the
coherence time $\tau_0$.  The critical temperature and the critical
pressure are estimated as $T_C\sim 12$ MeV and $P_C\sim 0.20$
MeV/fm$^3$, but these values may slightly depend on the choice of
$\tau_0$.  As for a future study, it is interesting to investigate the
dependence of the backbending and the critical point on the size and
the isospin composition of the system.

We have also shown constant volume caloric curves, but the signal of
the phase transition can not be seen clearly there.  It suggests that
it is important to identify the path of caloric curves drawn on
$E$-$V$ plain when those are constructed from experimental data to
discuss the existence of nuclear liquid-gas phase transition.

As we have demonstrated in this paper, AMD can be applied to ideal
thermal equilibrium situations as well as the dynamical processes of
the reactions. Therefore, it may be possible to construct the method
to extract the statistical information from experimental data by
studying the reaction simulation results of AMD together with the
calculations of ideal systems in thermal equilibrium.

\begin{acknowledgments}
This work was partly supported by High Energy Accelerator Research
Organization (KEK) as a supercomputer project.
\end{acknowledgments}

\appendix
\section{The modification of the energy for gaseous nucleons}
\label{ZEROPOINT}
The fragment number $\mathcal{N}_F$ utilized in this paper is similar
to the one in Ref.\ \cite{onoC12}, but a small modification is
introduced. The fragment number of Ref.\ \cite{onoC12} was defined as
\begin{equation}
\label{eq:NForg}
\mathcal{N}^{(0)}_F=\sum^A_{i=1}\mathcal{I}^{(0)}_i,\quad
\mathcal{I}^{(0)}_i=\frac{g(k_i)}{n_im_i},
\end{equation}
where
\begin{equation}
n_i=\sum_{j=1}^A\hat{f}_{ij},\; m_i=\sum_{j=1}^A\frac{1}{n_j}f_{ij},
\;k_i=\sum_{j=1}^A\bar{f}_{ij},
\end{equation}
and
\begin{equation}
\hat{f}_{ij}=F(d_{ij},\hat{\xi},\hat{a}),\;
f_{ij}=F(d_{ij},\xi,a),\;
\bar{f}_{ij}=F(d_{ij},\bar{\xi},\bar{a}),
\end{equation}
\begin{gather}
d_{ij}=|\text{Re}(\mathbf{Z}_i-\mathbf{Z}_j)|,\\
F(d,\xi,a)=\begin{cases}
1 & (d\leq a)\\
e^{-\xi(d-a)^2} & (d > a)
\end{cases},\\
g(k)=1+g_0e^{-(k-M)^2/2\sigma}.\\
\begin{tabular}{ccccccccc}
$\xi$  &$a$   &$\hat{\xi}$  &$\hat{a}$  &$\bar{\xi}$  &$\bar{a}$  
&$g_0$  &$\sigma$  &$M$  \\ \hline
2.0  &0.6 &2.0 &0.2 &1.0 &0.5 &1.0 &2.0 &12 \\ 
\end{tabular}
\end{gather}

In the present study, a gaseous nucleon is counted as a separate
`fragment' for the convenience of the temperature measurement.  We
allow a few gaseous nucleons get spatially close by regarding the
nucleon $i$ as gaseous when the number of the neighboring nucleons
including itself
\begin{equation}
q_i=\sum_{j=1}^Af_{ij}
\end{equation}
is less than about three.  This description is achieved by defining
the fragment number $\mathcal{N}_F$ as
\begin{equation}
\mathcal{N}_F=\sum_{i=1}^A \mathcal{I}_i,\quad
\mathcal{I}_i=(1-w(q_i))\mathcal{I}^{(0)}_i+w(q_i),
\label{eq:Ii}
\end{equation}
where
\begin{equation}
w(q)=F(q,1/2\sigma_q^2,Q)
\end{equation}
and the parameters are taken as $\sigma_q=0.85$ and $Q=2.5$.  The
newly introduced function $w(q_i)$ becomes zero unless $q\lesssim Q$
and the degree of isolation $\mathcal{I}_i$ reduces to the original one
$\mathcal{I}^{(0)}_i$. On the other hand, $\mathcal{I}_i$ becomes
close to one for the nucleons with $q_i\lesssim Q$.

The potential energies between the gaseous nucleons ($q_i\lesssim Q$)
are also modified in the present study. The term
$\mathcal{V}_{\text{mod}}$ of Eq.\ (\ref{eq:effectiveH}) is introduced
for this purpose.  The potential energy between two gaseous nucleons
is calculated by folding the widely spread wave functions with their
interaction.  The spread wave function, keeping the centroid position
in coordinate and momentum space same as the original Gaussian wave
function [Eq.\ (\ref{eq:AMDWF})], is given by
\begin{gather}
\phi_i(\mathbf{r};\alpha)=\left(\frac{2\nu}{\pi(1+\alpha)}\right)
^{\frac{3}{4}} \exp\left[-\frac{\nu}{1+\alpha}(\mathbf{r}
-\frac{\mathbf{Z}'_i}{\sqrt{\nu}})^2 \right]\label{eq:smearedWF}\\
\mathbf{Z}'_i=\text{Re}\mathbf{Z}_i+
i(1+\alpha)\text{Im}\mathbf{Z}_i,
\end{gather}
where $\alpha$ is a smearing parameter.  By denoting the potential
energy between nucleon $i$ and nucleon $j$ as $\mathcal{V}_{ij}$
calculated with the original Gaussian [Eq.\ (\ref{eq:AMDWF})] and
$\mathcal{V}_{ij}(\alpha)$ calculated with the smeared Gaussian [Eq.\
(\ref{eq:smearedWF})], $\mathcal{V}_{\text{mod}}$ is given by
\begin{equation}
\mathcal{V}_{\text{mod}}=\sum_{i<j}^A w(q_i)w(q_j)\Bigl[
-\mathcal{V}_{ij}+\mathcal{V}_{ij}(\alpha)\Bigr].
\end{equation}
The smearing parameter is chosen as $\alpha=3$ in order to reasonably
reproduce the binding energies of small nuclei such as deuteron, triton
and ${}^3\textrm{He}$ in case those stable compounds are formulated
during dynamical processes.  The density-dependent part of the
effective interaction is also included in $\mathcal{V}_{\text{mod}}$
with a similar approximation used in the triple-loop approximation
\cite{onoAu}. When we evaluate $V_{ij}(\alpha)$ with the
density-dependent part, we have also modified the density
$\rho(\mathbf{r})$ consistently with the transformation of the wave
function [Eq.\ (\ref{eq:smearedWF})] as
\begin{equation}
\rho(\mathbf{r};\alpha)=
\int\left(\frac{2\nu}{\pi\alpha}\right)^{\frac{3}{2}}
e^{-(2\nu/\alpha)(\mathbf{r}-\mathbf{r}')^2}
\rho(\mathbf{r}')d^3r'.
\end{equation}

\section{Temperature calculation by gaseous nucleons}
\label{ISOTEMP}
The microcanonical temperature is defined by $T^{-1}=(\partial
S/\partial E)|_{V,A}$.  We will show that this quantity can be
evaluated with a good precision by using the average kinetic energy of
a set of nucleons which can be chosen arbitrarily as long as two
conditions are satisfied. The first condition is that the quantum
effect should be negligible for any of these nucleons.  The
other condition is that these nucleons have to be chosen based on only
the nucleon spatial coordinates without using momentum variables. In
this section, we name the subsystem of these nucleons as
System 1 and the rest of the system as System 2.

The microcanonical ensemble $\{E,V,A\}$ can be divided into two
sub-ensembles: one with the number of nucleons of System 1 ($A_1$) is
larger than zero and the other with $A_1=0$. The density of
microstates of the total ensemble is given by the sum of the
microstates of these subensembles;
\begin{equation}
W(E)=W_{A_1>0}(E)+W_{A_1=0}(E).
\end{equation}
The constraints on the total volume $V$ and the total nucleon number
$A$ are omitted in this expression and the following for brevity. The
temperature of this ensemble is defined by
\begin{align}
\frac{1}{T}&=\frac{\partial \ln W(E)}{\partial E}\\
&=\frac{W_{A_1>0}(E)}{W(E)}\frac{1}{T_{A_1>0}}
+\frac{W_{A_1=0}(E)}{W(E)}\frac{1}{T_{A_1=0}},
\label{eq:Tdivide}
\end{align}
where $T_{A_1>0}\equiv(\partial(\ln W_{A_1>0}(E)/\partial E))^{-1}$
and $T_{A_1=0}\equiv(\partial(\ln W_{A_1=0}(E)/\partial E))^{-1}$ are
the microcanonical temperatures of the subensembles of $W_{A_1>0}$ and
$W_{A_1=0}$, respectively.  The contribution of the second term of Eq.\
(\ref{eq:Tdivide}) may be neglected if $W_{A_1=0}(E)$ is much less
than $W_{A_1>0}(E)$.

Let us denote the spatial coordinates of nucleons in System 1 by
\begin{equation}
R_1=\{\mathop{\text{Re}}\mathbf{Z}_i/\sqrt{\nu};\ i\in\mbox{System 1}\}.
\end{equation}
If the quantum effects are negligible in System 1, the density of
microstates of System 1 under the constraint of the nucleon positions
$R_1$ is given by
\begin{equation}
W_1(E_1,R_1)=\frac{(M/2\pi\hbar^2)^{\frac{3}{2}A_1}}
{A_1!\,\Gamma(\frac{3}{2}A_1)}
\Bigl(E_1-U(R_1)\Bigr)^{\frac{3}{2}A_1-1},
\label{eq:WforGAS}
\end{equation}
where $E_1$ is the energy of System 1 (excluding the interaction
energy with System 2), $U(R_1)$ is the potential energy within System
1 and $M$ is the nucleon mass.  The number of nucleons $A_1$ is
implicitly understood by $R_1$ in the left hand side and in the
following. Under the given condition $R_1$ of the nucleon positions of
System 1, we consider the density of microstates of System 2, which is
denoted by $W_2(E_2',R_1)$.  The nucleon positions of System 2 are
constrained by the condition that only the the nucleons of $R_1$
belong to System 1 when the defined algorithm is applied to the total
system. System 2 consists of $A-A_1$ nucleons.  The energy $E_2'$
includes the interaction energy between System 1 and System 2 in
addition to the internal energy of System 2 so that the total energy
is $E=E_1+E_2'$.  The explicit expression for $W_2(E_2',R_1)$ is not
necessary.  Using the density of microstates of System 1 and System 2,
$W_{A_1>0}(E)$ can be given by
\begin{equation}
W_{A_1>0}(E)
=\iint W_1(E-E_2',R_1)W_2(E_2',R_1)
dE_2'dR_1,
\label{eq:totalW}
\end{equation}
where the integral for $R_1$ also includes the summation over the
various cases of the nucleon number $A_1\;(>0)$ of System 1.  

Let us consider the case that $W_{A_1=0}(E)$ is negligible
compared with $W_{A_1>0}(E)$ and thus $W(E)$ is approximately equal to
$W_{A_1>0}(E)$. Then the microcanonical temperature $T$ can be
calculated by
\begin{align}
\frac{1}{T}\simeq&\frac{1}{T_{A_1>0}} 
=\frac{\partial\ln
W_{A_1>0}(E)}{\partial E}\notag\\
=&\frac{1}{W_{A_1>0}(E)}\iint \frac{\partial\ln
W_1(E-E_2',R_1)}{\partial E} \notag\\ 
&\times W_1(E-E_2',R_1)W_2(E_2',R_1)
dE_2'dR_1\notag\\
=& \biggl<\frac{\partial\ln
W_1(E_1,R_1)}{\partial E_1}\biggr>_{\{E,A_1>0\}}.
\label{eq:Twithlambda}
\end{align}
By inserting Eq.\ (\ref{eq:WforGAS}) to Eq.\ (\ref{eq:Twithlambda}),
we obtain
\begin{equation}
\frac{1}{T}\simeq\biggl<\frac{\frac{3}{2}A_1-1}{E_1-U(R_1)}\biggr>
_{\{E,A_1>0\}}
= \biggl<\frac{\frac{3}{2}A_1-1}{K_1}\biggr>_{\{E,A_1>0\}},
\label{eq:T1}
\end{equation}
where $K_1=E_1-U(R_1)$ is the total kinetic energy of System 1.  It is
also possible to evaluate the same temperature by
\begin{equation}
T\simeq\frac{2}{3}\biggl<\frac{K_1}{A_1}\biggr>_{\{E,A_1>0\}},
\label{eq:T2}
\end{equation}
if the nucleons in System 1 follow the Maxwell distribution, which is
usually the case with a good precision.  In fact, when the
distribution of the kinetic energy of System 1 is characterized by the
Boltzmann factor $e^{-K_1/T_0}$, Eq.\ (\ref{eq:T1}) can be calculated
as
\begin{align}
\left\langle\frac{\frac{3}{2}A_1-1}{K_1}
\right\rangle_{\{E,A_1>0\}}
&=\frac{\int
(\frac{3}{2}A_1-1)
K_1^{\frac{3}{2}A_1-2}e^{-K_1/T_0}
dK_1}
{\int K_1^{\frac{3}{2}A_1-1}
e^{-K_1/T_0}dK_1}\notag\\
&=\frac{(\frac{3}{2}A_1-1)T_0^{\frac{3}{2}A_1-1}\Gamma(\frac{3}{2}A_1-1)}
{T_0^{\frac{3}{2}A_1}\Gamma(\frac{3}{2}A_1)}\notag\\ 
&=T_0^{-1}.
\end{align}
On the other hand, Eq.\ (\ref{eq:T2}) can be calculated as
\begin{align}
\frac{2}{3}\left\langle\frac{K_1}{A_1}\right\rangle_{\{E,A_1>0\}}
&=\frac{2}{3A_1}\frac{\int K_1^{\frac{3}{2}A_1}
e^{-K_1/T_0}dK_1}
{\int K_1^{\frac{3}{2}A_1-1}
e^{-K_1/T_0}dK_1}\notag\\
&=\frac{2}{3A_1}\frac{T_0^{\frac{3}{2}A_1+1}\Gamma(\frac{3}{2}A_1+1)}
{T_0^{\frac{3}{2}A_1}\Gamma(\frac{3}{2}A_1)}\notag\\ 
&=T_0.
\end{align}

Equation (\ref{eq:T1}) is derived under the assumption that
$W_{A_1=0}(E)/W(E)$ is negligible. If $W_{A_1=0}(E)$ is not
negligible, the applicability of Eq.\ (\ref{eq:T1}) and Eq.\
(\ref{eq:T2}) is questioned unless $T_{A_1>0}\simeq T_{A_1=0}$ is
satisfied. If we can not evaluate $T_{A_1=0}$ practically, Eq.\
(\ref{eq:T1}) and Eq.\ (\ref{eq:T2}) should be used carefully when the
average number of the nucleons in System 1 is less than one.

\end{document}